\documentclass[aps,prd,twocolumn,showpacs,nofootinbib]{revtex4-1}

\usepackage{float}
\usepackage{graphicx}
\usepackage[colorlinks=true, linkcolor=blue, citecolor=blue, urlcolor=blue]{hyperref}

\begin{document}

\title{Hadronic production of $\Xi_{bc}$ with the intrinsic heavy-quark content at a fixed-target experiment at the LHC}

\author{Hong-Tai Li$^{a}$}
\email{liht@cqu.edu.cn}
\author{Xu-Chang Zheng$^{a}$}
\email{zhengxc@cqu.edu.cn}
\author{Jiang Yan$^{a}$}
\email{yjiang@cqu.edu.cn}
\author{Xing-Gang Wu$^{a}$}
\email{wuxg@cqu.edu.cn}
\author{Gu Chen$^{b}$}
\email{speecgu@gzhu.edu.cn}

\address{$^a$ Department of Physics, Chongqing Key Laboratory for Strongly Coupled Physics, Chongqing University, Chongqing 401331, China\\
$^b$ School of Physics and Electronic Engineering, Guangzhou University, Guangzhou 510006, China}

\begin{abstract}

In this paper, we make a detailed study on the hadronic production of the $\Xi_{bc}$ baryon at a fixed target experiment at the LHC (After@LHC). In estimating the production cross sections, the $(g+g)$, $(g+c)$ and $(g+b)$ production mechanisms are considered. For the initial heavy quarks, in addition to the extrinsic component, we also consider the intrinsic component. It is found that the $(g+c)$ and $(g+b)$ production mechanisms give sizable contributions to the $\Xi_{bc}$ production, and the $(g+b)$ mechanism dominates the production. The results show that there are about $3.40\times10^5$ $\Xi_{bc}$ events that can be produced per year at the After@LHC if the integrated luminosity of the After@LHC can be up to $2\,{\rm fb}^{-1}$ per year. Moreover, the intrinsic heavy quarks can have significant impact on the production, which inversely makes the intrinsic component be possibly tested at the After@LHC.


\end{abstract}

\maketitle

\section{Introduction}

In 2017, the doubly charmed baryon $\Xi_{cc}^{++}$ was first observed by the LHCb collaboration via the decay channel $\Xi_{cc}^{++} \to \Lambda_c^+ K^- \pi^+ \pi^+$ with $\Lambda_c^+ \to p K^- \pi^+$ \cite{Aaij:2017ueg}. This observation has been confirmed by the subsequent observation using the decay mode $\Xi_{cc}^{++} \to \Xi_{cc}^{+}\pi^+$ with $\Xi_{cc}^{+} \to pK^- \pi^+$ \cite{LHCb:2018pcs}. These observations on $\Xi_{cc}^{++}$ have aroused great interests in studying the doubly heavy baryons in the theoretical and experimental aspects. Compared to the heavy baryons that contain one heavy quark, the production of doubly heavy baryons involves more perturbative information which can be calculated perturbatively via proper QCD factorization approach. Therefore, the production of the doubly heavy baryons provides a good platform for studying QCD, especially the perturbative QCD (pQCD). In recent years, lots of theoretical works on the production of doubly heavy baryons at various high-energy colliders, e.g., $e^+e^-$, $ep$ and $pp$ (or $p\bar{p}$), have been carried out \cite{Jiang:2012jt, Jiang:2013ej, Chen:2014frw, Yang:2014ita, Yang:2014tca, Martynenko:2013eoa, Zheng:2015ixa, Bi:2017nzv, Sun:2020mvl, Chen:2014hqa, Chen:2019ykv, Chen:2018koh, Martynenko:2014ola, Koshkarev:2016acq, Koshkarev:2016rci, Groote:2017szb, Berezhnoy:2018bde, Brodsky:2017ntu, Berezhnoy:2018krl, Wu:2019gta, Qin:2020zlg, Niu:2018ycb, Niu:2019xuq, Zhang:2022jst, Luo:2022jxq, Luo:2022lcj, Ma:2022cgt} .

Up to now, only the doubly charmed baryon $\Xi_{cc}^{++}$ has been observed in experiments, while $\Xi_{bc}$~\footnote{For simplicity, throughout this paper, we use $\Xi_{QQ'}$ to denote the doubly heavy baryons that contain two heavy quarks $Q$, $Q'$ and a light quark. For instance, $\Xi_{bc}$ denotes the doubly heavy baryons $\Xi_{bc}^+$, $\Xi_{bc}^0$ and $\Omega_{bc}^0$.} and $\Xi_{bb}$ have not been observed in experiments~\footnote{The LHCb collaboration has carried out the searches for $\Xi_{bc}$, but it has not been observed to date \cite{LHCb:2020iko, LHCb:2021xba, LHCb:2022fbu}. In these searches, the exclusive decay channels $ {\Xi}_{bc}^0 \to D^{0}$pK$^{-}$, $\Xi^0_{bc} \to \Lambda^+_c \pi^-$, $\Xi^0_{bc} \to \Xi^+_c \pi^-$ and $\Xi_{bc}^{+} \to J/\it{\psi} \Xi_{c}^{+}$ were used to reconstruct $\Xi_{bc}$. Recently, an inclusive approach through the decay $\Xi_{bc}\to \Xi_{cc}^{++}+X$ was proposed for searching $\Xi_{bc}$ \cite{Qin:2021zqx}, which provides a new opportunity for searching $\Xi_{bc}$ at the LHCb.}. The observation of $\Xi_{bc}$ and $\Xi_{bb}$ is attractive and can help to further understand the strong interaction. A fixed-target experiment (AFTER@LHC) using the LHC proton and heavy-ion beams has been proposed to probe the high-$x$, spin and quark gluon plasma (QGP) physics \cite{Lansberg:2012wj, Lansberg:2013wpx, Brodsky:2012vg, Barschel:2020drr}. The center-of-mass energy per nucleon-nucleon collision ($\sqrt{s_{NN}}$) of the AFTER@LHC can reach up to $115\,{\rm GeV}$ for $pp/pA$ collisions, and the luminosity of the AFTER@LHC will be very high due to the large density of the target. With the high energy and high luminosity, it has been pointed out that the AFTER@LHC will become a good platform for studying the properties of $\Xi_{cc}$~\cite{Chen:2014hqa, Chen:2019ykv}. As a rough order estimation, sizable $\Xi_{bc}$ baryons may also be produced at the AFTER@LHC. Thus in the present paper, we will devote ourselves to the production of $\Xi_{bc}$ at the AFTER@LHC.

The existing calculations for the hadronic production of $\Xi_{bc}$ are mainly based on the gluon-gluon fusion ($g+g$)  mechanism \cite{Baranov:1995rc, Zhang:2011hi, Berezhnoy:2018bde}. The authors of Ref.\cite{Chang:2005wd} found that in some kinematic regions, e.g. the small transverse momentum ($p_t$) region, the gluon-heavy-quark collisions ($g+c$) and ($g+b$) can give important contributions to the hadronic production of the $B_c(B_c^*)$ meson, which even exceed the contribution of the ($g+g$) mechanism. The production mechanism of the $\Xi_{bc}$ baryon is similar to that of the $B_c(B_c^*)$ meson. Moreover, the measured $p_t$ of doubly heavy baryons could be very small at a fixed-target experiment such as the AFTER@LHC. Hence, it is expected that the ($g+c$) and ($g+b$) mechanisms can also give sizable contributions to the hadronic production of $\Xi_{bc}$ at the AFTER@LHC. In this work, in addition to the ($g+g$) mechanism, we will also consider the contributions of the ($g+c$) and ($g+b$) mechanisms.

The heavy quarks in a nucleon have either perturbative ``extrinsic" or nonperturbative ``intrinsic" origins~\cite{Brodsky:1980pb,Brodsky:1981se,Brodsky:2015fna}. The extrinsic heavy quarks are generated by gluon splitting in the DGLAP evolution, while the intrinsic heavy quarks are non-perturbative and arise from the wave function of the nucleon, which even exist for scales below the heavy quark threshold. Since the ($g+c$) and ($g+b$) mechanisms are expected to give important contributions to the $\Xi_{bc}$ production at the AFTER@LHC, the cross sections of $\Xi_{bc}$ may be sensitive to the parton distribution functions (PDFs) of the heavy quarks. Therefore, it is interesting to study the impact of the intrinsic heavy quarks on the $\Xi_{bc}$ production.

The paper is organized as follows. In Sec.II, we present formulas for calculating the hadronic production of $\Xi_{bc}$. In Sec.III, we present the numerical results and the related discussions. The final section is reserved for a summary.

\section{Calculation Technology}
	
According to the pQCD factorization and the GM-VFN scheme \cite{Aivazis:1993kh,Aivazis:1993pi,Olness:1997yc,Amundson:2000vg,Amundson:2000vg}, the cross section for the hadronic production of $\Xi_{bc}$ via the collision of two protons can be written as
\begin{widetext}
\begin{eqnarray}
\sigma(H_1+H_2 \to \Xi_{bc}+X) =&& f^{g}_{H_{1}}(x_{1},\mu) f^{g}_{H_{2}}(x_{2},\mu) \otimes \hat{\sigma}_{gg \rightarrow\Xi_{bc}}(x_{1},x_{2},\mu)\nonumber \\
&& + \sum_{i,j=1,2;i\neq j} f^{g}_{H_{i}} (x_{1},\mu)\left[f^{c}_{H_{j}}(x_{2},\mu) - f^{c}_{H_{j}}(x_{2},\mu)_{\rm SUB} \right] \otimes \hat{\sigma}_{gc\rightarrow \Xi_{bc}}(x_{1},x_{2},\mu)\nonumber \\
&& + \sum_{i,j=1,2;i\neq j} f^{g}_{H_{i}} (x_{1},\mu)\left[f^{b}_{H_{j}}(x_{2},\mu) - f^{b}_{H_{j}}(x_{2},\mu)_{\rm SUB} \right] \otimes \hat{\sigma}_{gb\rightarrow \Xi_{bc}}(x_{1},x_{2},\mu)+ \cdots ,
\end{eqnarray}
\end{widetext}
where $f^{i}_{H}(x,\mu)$ (with $H=H_1$ or $H_2$; $x=x_1$ or $x_2$; $i=g$, $c$ or $b$) is the PDF of the parton $i$ in the incident proton $H$. We have implicitly set the factorization scale and the renormalization scale to be the same, i.e. $\mu_F=\mu_R=\mu$. $f^{Q}_{H}(x,\mu)_{\rm SUB}$ (with $Q=c,b$) is the subtraction term, which is introduced to avoid double counting between the $(g+g)$ and $(g+Q)$ contributions. The subtraction term~\cite{Aivazis:1993kh,Aivazis:1993pi,Olness:1997yc,Amundson:2000vg} is defined as
\begin{eqnarray}
f^{Q}_{H}(x,\mu)_{\rm SUB}& \equiv & f^{g}_{H}(x,\mu) \otimes f^{Q}_g(x,\mu) \nonumber\\
&=& \int^1_{x}\frac{dy}{y}f^{Q}_g(y,\mu) f^{g}_{H}\left(\frac{x}{y},\mu\right)
\label{subtraction}
\end{eqnarray}
with
\begin{eqnarray}
f^Q_g(x,\mu) &=& \frac{\alpha_s(\mu)}{2\pi} \ln\frac{\mu^2}{m^2_Q}P_{g\to Q}(x) \nonumber\\
&=&\frac{\alpha_s(\mu)} {4\pi}(1-2x+2x^2)\ln\frac{\mu^2}{m^2_Q}.
\label{subtraction2}
\end{eqnarray}

According to the NRQCD factorization \cite{Bodwin:1994jh}, the cross section $\hat\sigma_{gi\rightarrow \Xi_{bc}}$ (with $i=g,b,c$) can be further factorized as \cite{Ma:2003zk,Chang:2006eu,Chang:2006xp}
\begin{eqnarray}
\hat{\sigma}_{gi\rightarrow \Xi_{bc}}=&&  H(gi\to (bc)_{\bf\bar 3}[^1S_0] ) \cdot h^{(bc)}_{\bar 3} \nonumber\\
&& +H(gi\to (bc)_{\bf 6}[^1S_0])\cdot h^{(bc)}_6 \nonumber\\
&& +H(gi\to (bc)_{\bf\bar 3}[^3S_1] ) \cdot {h'}^{(bc)}_{\bar 3}  \nonumber\\
&& +H(gi\to (bc)_{\bf 6}[^3S_1] ) \cdot {h'}^{(bc)}_6 +\cdots,
\end{eqnarray}
where the ellipsis denotes the higher-order terms in $v$, and $v$ is the relative velocity between the $b$ quark and the $c$ quark in the rest frame of $\Xi_{bc}$. $H(gi\to (bc)_{\bf\bar{3},6}[^1S_0])$ and $H(gi\to (bc)_{\bf\bar{3},6}[^3S_1] )$ are the short-distance coefficients (SDCs), which describe the production of the free $(bc)$ quark pair with proper (color and spin) quantum numbers. Since the involved energy scales are larger than the heavy quark threshold, the SDCs can be calculated within pQCD. The long-distance matrix elements (LDMEs) $h^{(bc)}_{\bar{3}}$, $h^{(bc)}_6$, ${h'}^{(bc)}_{\bar{3}}$ and ${h'}^{(bc)}_6$ describe the transitions of the $(bc)$ quark pair in $[^1S_0]_{\bf \bar 3}$, $[^1S_0]_{\bf 6}$, $[^3S_1]_{\bf \bar 3}$, $[^3S_1]_{\bf 6}$ spin and color configurations into the $\Xi_{bc}$ baryon, respectively. These LDMEs are nonperturbative in nature, so they cannot be calculated within pQCD. $h^{(bc)}_3$ and $h'^{(bc)}_3$ can be related to the wavefunction of the color ${\bf\bar{3}}$ diquark \footnote{The transition from the free $(bc)$ pair in color ${\bf\bar{3}}$ state to $\Xi_{bc}$ can be regarded as two steps: the formation of a bounded diquark in color ${\bf\bar{3}}$, and the fragmentation from the diquark into $\Xi_{bc}$. The probability of the first step can be described through the wavefunction of the color ${\bf\bar{3}}$ diquark. Since the color ${\bf\bar{3}}$ diquark can pick up a light quark from the collision environment easily, the probability of the fragmentation from the color ${\bf\bar{3}}$ diquark into $\Xi_{bc}$ is usually assumed as $\sim 100\%$. Thus, the total probability of the two steps can be described by the wavefunction of the color ${\bf\bar{3}}$ diquark.}, i.e. $h'^{(bc)}_3\simeq h^{(bc)}_3\simeq\vert\Psi_{bc}(0)\vert^2$. However, the color ${\bf 6}$ LDMEs $h^{(bc)}_6$ and ${h'}^{(bc)}_6$ do not have such a relation. According to the discussions presented in Ref.\cite{Ma:2003zk}, the color ${\bf 6}$ LDMEs $h^{(bc)}_6$ and ${h'}^{(bc)}_6$ are of the same order in $v$ as the color ${\bf\bar{3}}$ LDMEs $h^{(bc)}_{\bar{3}}$ and $h'^{(bc)}_{\bar{3}}$, thus we take them all as $\vert \Psi_{bc}(0)\vert^2$ to make our estimates. Due to the fact that all these LDMEs are overall parameters in the calculations, we can easily improve our numerical results when we have more accurate values of these LDMEs.

\begin{figure}[htb]
\includegraphics[width=0.5\textwidth]{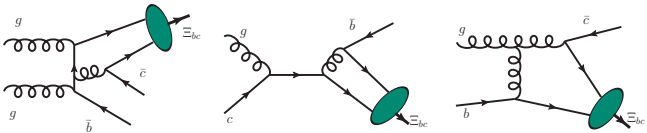}
\caption{Typical Feynman diagrams for the subprocesses of the hadronic production of $\Xi_{bc}$.}
\label{feynman}
\end{figure}

There are 36 Feynman diagrams for the $(g+g)$ mechanism, and 5 Feynman diagrams for the $(g+c)$ or $(g+b)$ mechanism. Three of these Feynman diagrams are shown in Fig.\ref{feynman}. From the figure, we can see that the cross section for the $(g+g)$ mechanism starts at order $\alpha_s^4$, while the cross sections for the $(g+c)$ and $(g+b)$ mechanisms start at order $\alpha_s^3$. Compared to the $(g+g)$ mechanism, the cross sections for the $(g+c)$ and $(g+b)$ mechanisms are enhanced by a factor of $1/\alpha_s$ in the short-distance part. Therefore, although the PDFs for the charm and bottom quarks are highly suppressed compared to the gluon PDF, the $(g+c)$ and $(g+b)$ mechanisms may give important contributions to the $\Xi_{bc}$ production in some kinematic regions.

Considering the intrinsic heavy-quark component, the PDF $f_H^i$ ($i=g,c,b$) can be expressed as
\begin{eqnarray}
f_H^i(x,\mu)&=& f_H^{i,0}(x,\mu)+f_H^{i,\rm in}(x,\mu),
\label{apsg}
\end{eqnarray}
where $f_H^{i,0}$ is the PDF without the intrinsic heavy-quark effect, and $f_H^{i,\rm in}(x,\mu)$ is the term due to the intrinsic heavy-quark effect. It is noted that the gluon PDF at a scale larger than heavy quark threshold is affected by the intrinsic heavy-quark effect through the DGLAP evolution, thus the gluon PDF also contains two terms as shown in Eq.(\ref{apsg}). The PDFs without the intrinsic heavy-quark effect have been determined by several groups through global fitting of experimental data.

For the intrinsic charm, several models are proposed to estimate its distribution, e.g., the BHPS model \cite{Brodsky:1980pb}, the meson cloud models \cite{Navarra:1995rq,Hobbs:2013bia}, the sealike model \cite{Pumplin:2005yf}, etc. In this work, we adopt the BHPS model for $f^{c,\rm in}_H(x,2m_c)$ to discuss the effect of the intrinsic charm, i.e.,
\begin{eqnarray}
&&f^{c,\rm in}_H(x,2m_c)   \nonumber\\
&&= 6  x^2 \xi \left[6x(1+x) \ln x + (1-x) (1 + 10x + x^2) \right],
\label{bhps}
\end{eqnarray}
where the factor $\xi$ depends on the probability of finding the intrinsic charm quark, i.e.,
\begin{displaymath}
A_{\rm in}\equiv \int_0^1 f^{c, \rm in}_H(x,2m_c)\; dx=\xi \times 1\% \;.
\end{displaymath}
The probability for finding the intrinsic $c/\bar{c}$-component in the proton at the fixed low-energy scale $2m_c$ is estimated to be about $1\%$~\cite{Brodsky:1980pb, Brodsky:1981se}, and we take a range of $\xi\in[0.1,1]$ for discussion\footnote{By using the ATLAS data, the authors of Ref.\cite{Bednyakov:2017vck} have presented an upper limit to the intrinsic charm probability, $A_{\rm in}<1.93\%$. The range of $A_{\rm in}$ we adopted is compatible with this upper limit.}.

In the BHPS model, the heavy quark mass is not explicit in the distribution function, it only affects the probability for the heavy quark, i.e. $\xi$. Hence, the distribution function for the intrinsic charm given in Eq.(\ref{bhps}) also applies to the intrinsic bottom, except that we should change the value of $\xi$ for the intrinsic bottom properly. It is expected that the probability for the intrinsic bottom is suppressed with respect to the intrinsic charm by a factor $m_c^2/m_b^2 \sim 0.1$. Thus, we estimate the distribution function for the intrinsic bottom through the following relation~\cite{Lyonnet:2015dca}
\begin{eqnarray}
f_H^{b,\rm in}(x,2m_c)=\frac{m^2_c}{m^2_b}f_H^{c,\rm in}(x,2m_c).
\end{eqnarray}

With the PDFs for the intrinsic heavy quarks at the initial scale $2 m_c$, we can obtain the PDFs at any other scale by solving the DGLAP equations. We adopt the approximate method introduced in Ref.\cite{Field:1989uq} to solve the DGLAP equations, and we obtain
\begin{widetext}
\begin{eqnarray}
f^{Q,\rm in}_H(x,\mu)=&& \int_x^1 \frac{dy}{y}  \left\{f^{Q,\rm in}_H(x/y,2m_c) \frac{[-\ln(y)]^{a_Q\kappa-1}}{\Gamma(a_Q \kappa)}\right\} \nonumber\\
&& +\kappa\int_x^1\frac{dy}{y} \int_y^1 \frac{dz}{z} \left \{f^{Q,\rm in}_H(y/z,2m_c) \frac{[-\ln(z)]^{a_Q\kappa-1}}{\Gamma(a_Q\kappa)} P_{\Delta Q}(x/y) \right \}+{\cal O}(\kappa^2),\label{intrc} \\
f^{g,\rm in}_H(x,\mu)=&&\sum_{Q=c,b} \frac{2\kappa}{a_g-a_Q}\int_x^1 \frac{dy}{y}\int_{a_Q}^{a_g}da\int_y^1 \frac{dz}{z}\Big\{\frac{[-\ln(z)]^{a\kappa-1}}{\Gamma(a\kappa)} f^{Q,\rm in}_H(z,2m_c) P_{Q \to gQ}(x/y)\Big\}+{\cal O }(\kappa^2),
\label{intrg}
\end{eqnarray}
\end{widetext}
where $Q=c$ or $b$, and
\begin{eqnarray}
&&a_g = 6, a_{Q}=\frac{8}{3}, \beta_0=11-2n_f/3, \nonumber\\
&&\kappa=\frac{2}{\beta_0} \ln \left[\frac{\alpha_s(2m_c)}{\alpha_s(\mu)}\right], \nonumber\\
&&P_{\Delta Q}(x) = \frac{4}{3} \left[\frac{1+x^2}{1-x}+\frac{2}{\ln x}+\left(\frac{3}{2}-2\gamma_E\right)\delta(1-x)\right], \nonumber\\
&& P_{Q\to gQ} = \frac{4}{3}\left[\frac{1+(1-x)^2}{x}\right].
\label{intrg1}
\end{eqnarray}

In the calculation, we shall use GENXICC~\cite{Chang:2007pp, Chang:2009va, Wang:2012vj} to simulate the hadronic production of $\Xi_{bc}$. GENXICC is an effective generator for simulating the production of the doubly heavy baryons ($\Xi_{cc}$, $\Xi_{bc}$ and $\Xi_{bb}$), and has been widely used in experiments~\cite{Traill:2018hhc, LHCb:2018zpl, LHCb:2018pcs, LHCb:2019gqy, LHCb:2019qed, LHCb:2021eaf, LHCb:2021xba, LHCb:2020iko}. Here, we modify the generator properly so as to include the contributions from both extrinsic and intrinsic heavy quarks.

\section{Numerical results and discussions}
\label{results}

For the numerical calculation, the input parameters are taken as follows:
\begin{eqnarray}
&&m_c=1.8\,{\rm GeV},~~~~~ m_b=5.1\,{\rm GeV},\nonumber \\
&&M_{\Xi_{bc}}=6.9\,{\rm GeV},\; \vert \Psi_{bc}(0)\vert^2 =0.065\,{\rm GeV}^3,
\end{eqnarray}
which are the same as those used in Ref.\cite{Baranov:1995rc}. The factorization and the renormalization scales are taken as the ``transverse mass" of $\Xi_{bc}$, i.e. $\mu=m_T=\sqrt{p_t^2+M_{\Xi_{bc}}^2}$. For the extrinsic PDFs, we adopt the CT14LO PDF~\cite{ct14lo} version extracted by the CTEQ group. For the probability ($A_{\rm in}$) of the intrinsic charm quark in the proton, we take several values, i.e., $A_{\rm in}=0, 0.1\%, 0.3\%$, and $1\%$ to see the effects of the intrinsic heavy-quarks, where $A_{\rm in}=0$ corresponds to the case that only the extrinsic mechanism is considered for the heavy quarks in the proton. We implicitly take a small transverse momentum cut for the $\Xi_{bc}$ events, i.e. $p_t > 0.2$GeV, which was used in the SELEX experiment \cite{SELEX:2002wqn} and could be adopted by the fixed-target experiment at the After@LHC.

In order to have an overall impression on the hadronic production of $\Xi_{bc}$ at the After@LHC, we first present the integrated cross sections for the $\Xi_{bc}$ production via the $(g+g)$, $(g+b)$ and $(g+c)$ mechanisms in Table \ref{tb.section}, where the contributions for different intermediate $(bc)$ states, i.e., $(bc)_{\bf\bar{3}}[^1S_0]$, $(bc)_{\bf 6}[^1S_0]$, $(bc)_{\bf\bar{3}}[^3S_1]$ and $(bc)_{\bf 6}[^3S_1]$ are shown explicitly. From Table \ref{tb.section}, we can see that in addition to the color ${\bf\bar{3}}$ intermediate states, the color ${\bf 6}$ intermediate states also give important contributions. For the $(g+g)$ and $(g+c)$ mechanisms, the contribution from the color ${\bf 6}$ intermediate states is even larger than that from  the color ${\bf\bar{3}}$ intermediate states. Comparing the contributions from different mechanisms, i.e., $(g+g)$, $(g+c)$ and $(g+b)$, we can see that the $(g+c)$ and $(g+b)$ mechanisms give sizable contributions, and the $(g+b)$ mechanism dominates over the $\Xi_{bc}$ production. The results also show that the intrinsic heavy quarks can have significant impact on the cross sections of the $(g+c)$ and $(g+b)$ mechanisms. For example, even if there is only $0.1\%$ probability to find the intrinsic charm-quark component in the proton, i.e. $A_{\rm in}=0.1\%$, the cross sections for the $(g+c)$ and $(g+b)$ mechanisms are increased by about $8.0\%$ and $11.4\%$, respectively.

\begin{widetext}
\begin{center}
\begin{table}[htb]
\begin{tabular}{|c|ccc|ccc|ccc|ccc|}
\hline
~~~-~~~ & \multicolumn{3}{c}{$A_{\rm in}=0$} & \multicolumn{3}{|c|} {$A_{\rm in}=0.1\%$} & \multicolumn{3}{|c|} {$A_{\rm in}=0.3\%$} & \multicolumn{3}{|c|} {$A_{\rm in}=1\%$} \\
\hline
- & $\sigma_{g+g}$ & $\sigma_{g+c}$ & $\sigma_{g+b}$ & $\sigma_{g+g}$ & $\sigma_{g+c}$ & $\sigma_{g+b}$ & $\sigma_{g+g}$ & $\sigma_{g+c}$ & $\sigma_{g+b}$& $\sigma_{g+g}$ & $\sigma_{g+c}$ & $\sigma_{g+b}$   \\
\hline
$(bc)_{\bar{{\bf 3}}}[^1S_0]$ &~1.45 &~0.86 &~4.67 &~1.45 &~0.93 &~5.21 &~1.46 &~1.08 &~6.27 &~1.48 &~1.59 &~10.03   \\
$(bc)_{{\bf 6}}[^1S_0]$       &~1.35 &~0.70 &~4.22 &~1.35 &~0.76 &~4.70 &~1.36 &~0.87 &~5.64 &~1.37 &~1.27 &~~8.97   \\
$(bc)_{\bar{{\bf 3}}}[^3S_1]$ &~5.18 &~5.06 &28.26 &~5.19 &~5.47 &31.49 &~5.21 &~6.28 &37.96 &~5.29 &~9.14 &~60.51   \\
$(bc)_{{\bf 6}}[^3S_1]$       &~9.48 &~5.97 &23.53 &~9.49 &~6.44 &26.20 &~9.53 &~7.38 &31.53 &~9.67 &10.67 &~50.17   \\
Total                         &17.46 &12.59 &60.68 &17.48 &13.60 &67.60 &17.56 &15.61 &81.40 &17.81 &22.67 &129.68   \\
\hline
\end{tabular}
\caption{Cross sections (in unit pb) for the $\Xi_{bc}$ production at the After@LHC with different intrinsic heavy-quark components corresponding to different choices of $A_{\rm in}$, i.e., $A_{\rm in}=0$, $0.1\%$, $0.3\%$, and $1\%$, respectively. Here, the transverse momentum cut has been taken as $p_t>0.2\;\rm GeV$.}
\label{tb.section}
\end{table}
\end{center}
\end{widetext}

If the integrated luminosity of the After@LHC reaches $0.05\,{\rm fb}^{-1}$ or $2\,{\rm fb}^{-1}$ per operation year~\cite{Brodsky:2012vg}, there are about $4.54\times10^3$ or $1.81\times10^5$ $\Xi_{bc}$ events to be generated at the After@LHC for $A_{\rm in}=0$. If setting $A_{\rm in}=1\%$, the $\Xi_{bc}$ events shall be greatly increased to $8.51\times10^3$ or $3.40\times10^5$ per operation year. The number of the $\Xi_{bc}$ events is sensitive to the probability of finding intrinsic heavy quarks in the proton. Therefore, the $\Xi_{bc}$ production at the After@LHC provides a good platform for testing and studying the intrinsic heavy-quark content.

\subsection{$\Xi_{bc}$ production via the $(g+g)$ mechanism}

In this subsection we shall analyze the cross sections for the $\Xi_{bc}$ production via the $(g+g)$ mechanism.

\begin{figure}[htb]
\includegraphics[width=0.45\textwidth]{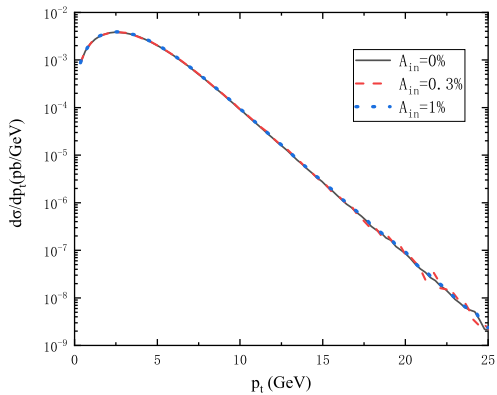}
\caption{Comparison of the $p_t$ distributions for the hadronic production of $\Xi_{bc}$ with and without the intrinsic heavy quarks, $A_{\rm in}=0$, $A_{\rm in}=0.3\%$ and $A_{\rm in}=1\%$, via the $(g+g)$ mechanism at the After@LHC, where the contributions from various intermediate $(bc)$ states have been summed up.}  \label{ptgg1}
\end{figure}

\begin{figure}[htb]
\includegraphics[width=0.45\textwidth]{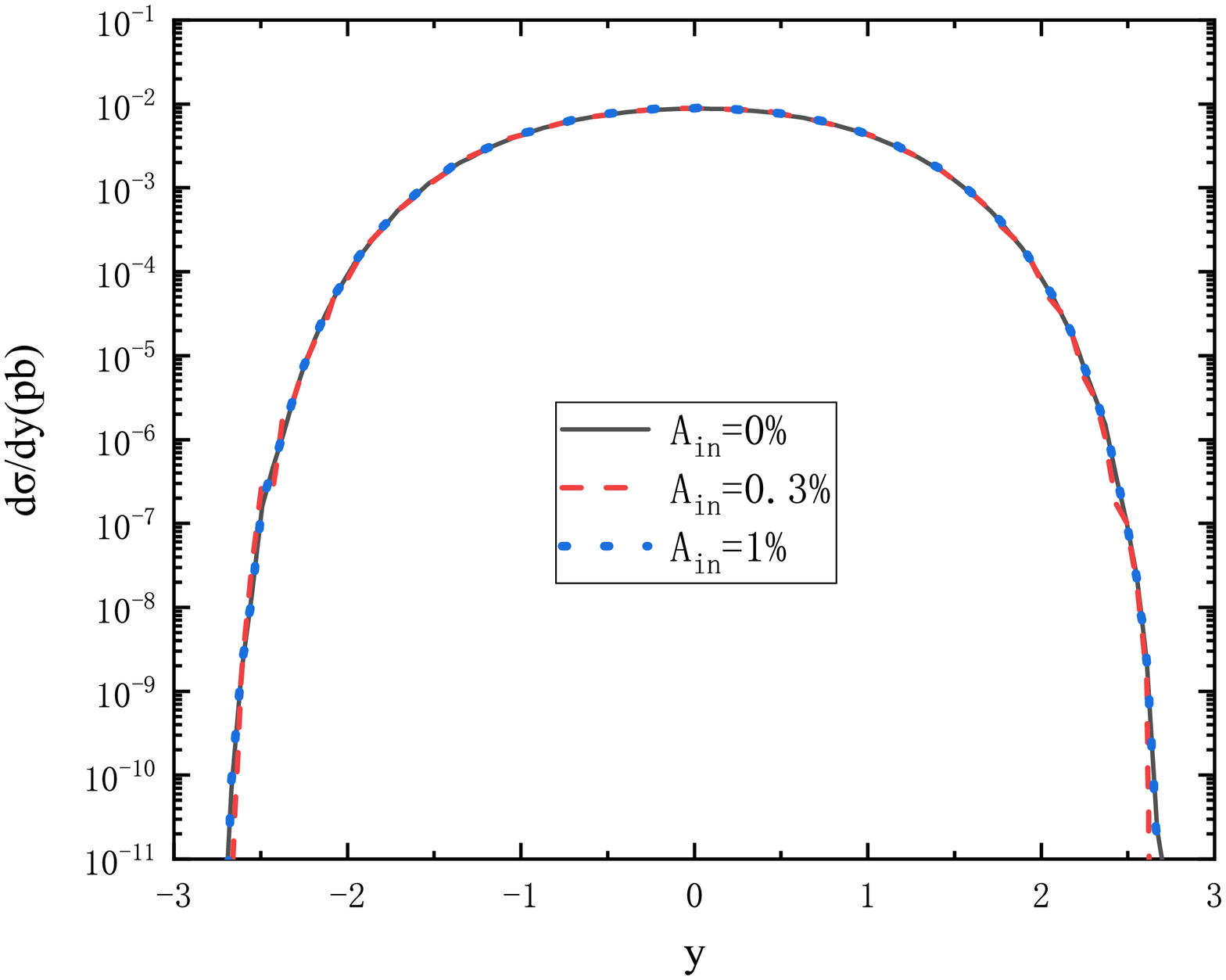}
\caption{Comparison of the $y$ distributions for the hadronic production of $\Xi_{bc}$ with and without the intrinsic heavy quarks, $A_{\rm in}=0$, $A_{\rm in}=0.3\%$ and $A_{\rm in}=1\%$, via the $(g+g)$ mechanism at the After@LHC, where the contributions from various intermediate $(bc)$ states have been summed up. Here, the transverse momentum cut is taken as $p_t>0.2\;\rm GeV$.}
\label{rapgg1}
\end{figure}

\begin{figure}[htb]
\includegraphics[width=0.45\textwidth]{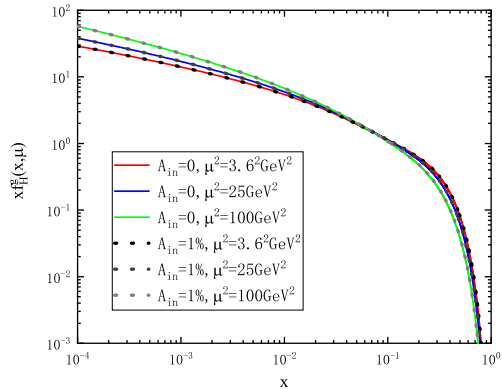}
\caption{The gluon PDF with and without intrinsic heavy quarks, $A_{\rm in}=1\%$ and $A_{\rm in}=0$, at different scales.}
\label{pdfg}
\end{figure}

\begin{figure}[htb]
\includegraphics[width=0.5\textwidth]{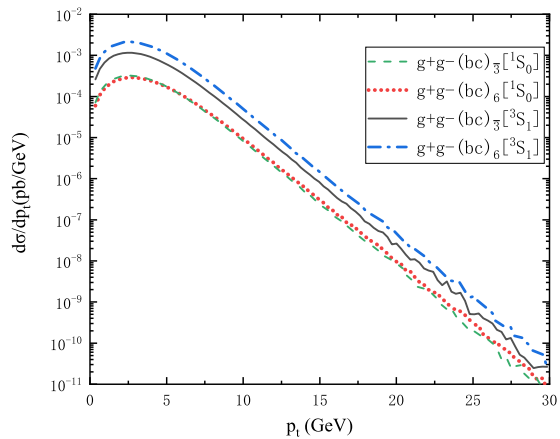}
\caption{The $p_t$ distributions of the $\Xi_{bc}$ production via the $(g+g)$ mechanism with $A_{\rm in}=1\%$ at the After@LHC, in which no $y$ cut has been applied. The contributions from different intermediate $(bc)$ states are presented explicitly.}  \label{ptggtu}
\end{figure}

\begin{figure}[htb]
\includegraphics[width=0.45\textwidth]{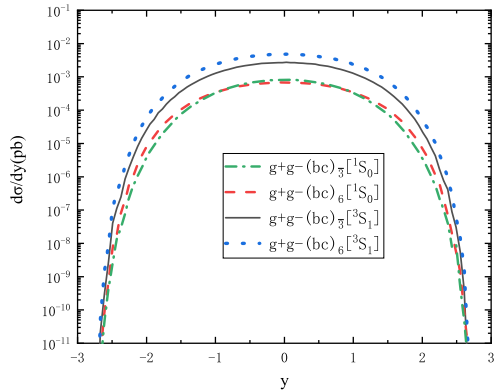}
\caption{The $y$ distributions of the $\Xi_{bc}$ production via the $(g+g)$ mechanism with $A_{\rm in}=1\%$ at the After@LHC, in which $p_t>0.2\;\rm GeV$ has been taken. The contributions from different intermediate $(bc)$ states are presented explicitly.}  \label{rapggtu}
\end{figure}

In order to see how the differential cross sections of $\Xi_{bc}$ via the $(g+g)$ mechanism are affected by the intrinsic heavy quarks, we present the differential $p_t$ (transverse momentum) and $y$ (rapidity) distributions for the $\Xi_{bc}$ production via the $(g+g)$ mechanism with different intrinsic heavy-quark probabilities in Figs.~\ref{ptgg1} and~\ref{rapgg1}. Figs.~\ref{ptgg1} and~\ref{rapgg1} show that the $p_t$ and $y$ distributions of $\Xi_{bc}$ change very slightly in whole $p_t$ or $y$ region after including the intrinsic heavy-quark content. This is because the effect of the intrinsic heavy quarks on the gluon PDF is small. To show this point more obviously, a comparison of the gluon PDF with and without intrinsic heavy-quark effect is presented in Fig.~\ref{pdfg}, where three typical scales, i.e., $\mu=3.6\;{\rm GeV}$, $5\;{\rm GeV}$ and $10\;{\rm GeV}$, are adopted. The two curves with and without intrinsic heavy quarks almost coincide under various scales, indicating the intrinsic heavy-quark effect on the gluon PDF is negligible.

It is interesting to see the differential distributions of the contributions from different intermediate $(bc)$ states. In Figs.\ref{ptggtu} and \ref{rapggtu}, the differential $p_t$ and $y$ distributions for different intermediate states, i.e., $(bc)_{\bar{{\bf 3}}}[^1S_0]$, $(bc)_{{\bf 6}}[^1S_0]$, $(bc)_{\bar{{\bf 3}}}[^3S_1]$ and $(bc)_{{\bf 6}}[^3S_1]$ are presented, where the probability of the intrinsic charm is fixed as $A_{\rm in}=1\%$. The figures show that the curves for the contributions of different intermediate $(bc)$ states are very similar in shapes. The $(bc)_{{\bf 6}}[^3S_1]$ channel dominates the production via the $(g+g)$ mechanism.

\subsection{$\Xi_{bc}$ production via the $(g+c)$ and $(g+b)$ mechanisms}

In this subsection we shall analyze the cross sections for the $\Xi_{bc}$ production via the $(g+c)$ and $(g+b)$ mechanisms.

\begin{figure}[htb]
\includegraphics[width=0.5\textwidth]{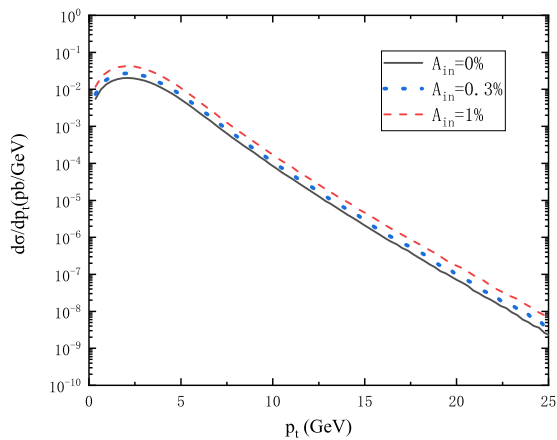}
\caption{Comparison of the $p_t$ distributions for the $\Xi_{bc}$ production with and without the intrinsic heavy quarks, $A_{\rm in}=1\%$, $A_{\rm in}=0.3\%$ and $A_{\rm in}=0$, via the $(g+c)$ and $(g+b)$ mechanisms at the After@LHC. Here, the contributions from various intermediate $(bc)$ states have been summed up.}
\label{ptgc}
\end{figure}

\begin{figure}[htb]
\includegraphics[width=0.5\textwidth]{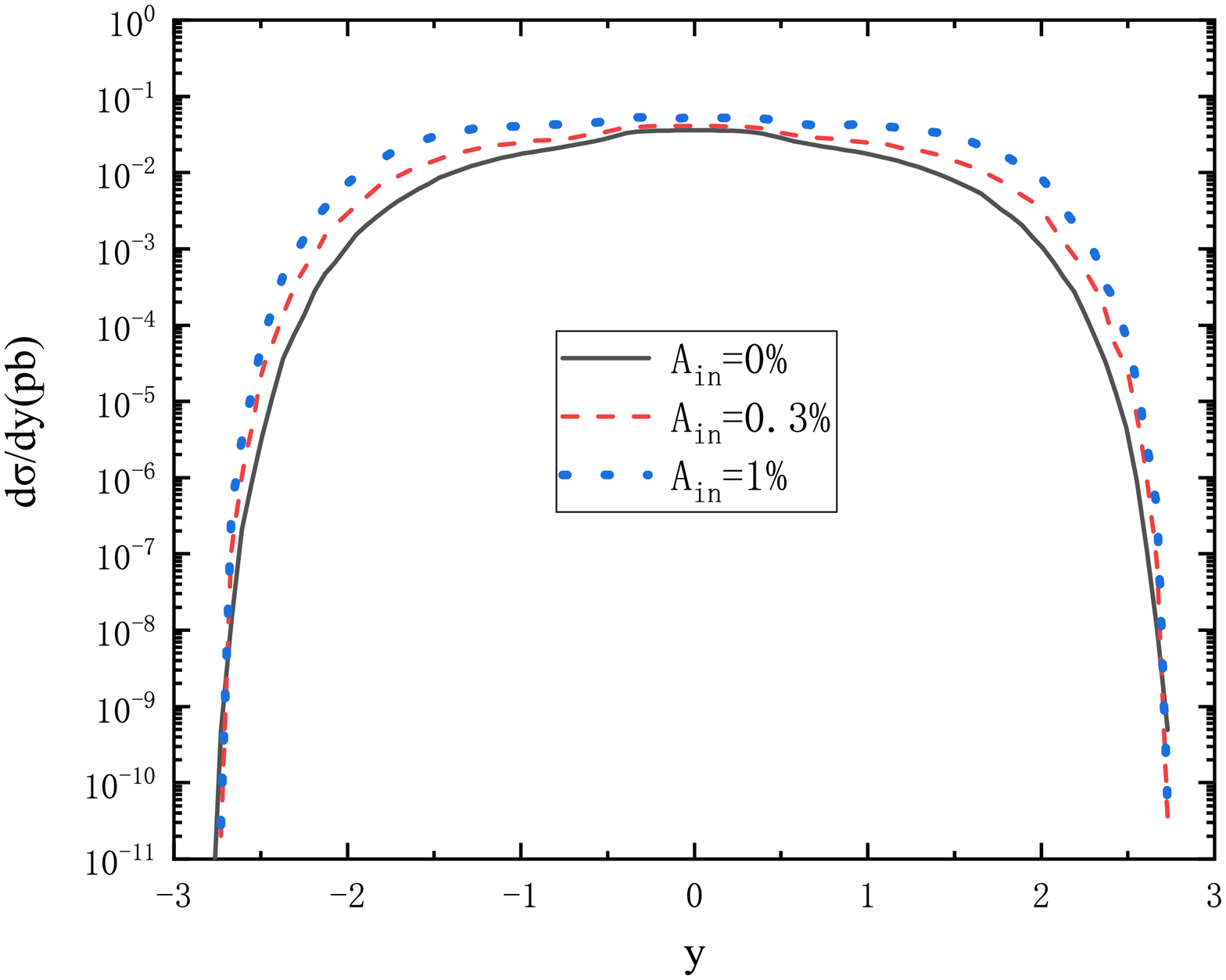}
\caption{Comparison of the $y$ distributions for the $\Xi_{bc}$ production with and without intrinsic heavy quarks, $A_{\rm in}=1\%$, $A_{\rm in}=0.3\%$ and $A_{\rm in}=0$, via the $(g+c)$ and $(g+b)$ mechanisms at the After@LHC, in which $p_t>0.2\;\rm GeV$ has been taken. The contributions from various intermediate $(bc)$ states have been summed up.}
\label{rapgc}
\end{figure}

\begin{figure}[htb]
\includegraphics[width=0.5\textwidth]{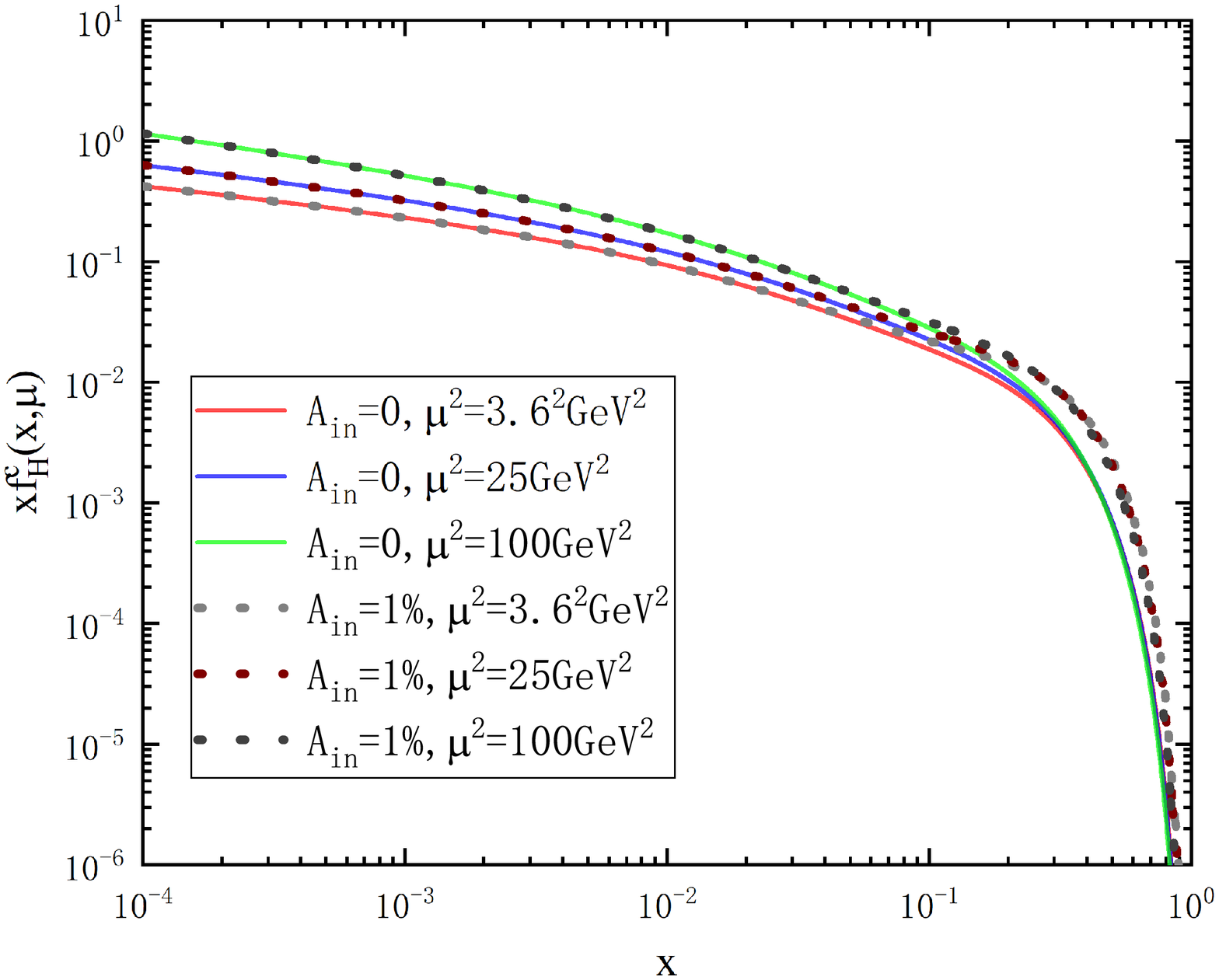}
\caption{The charm and bottom PDFs with and without intrinsic heavy-quark component, $A_{\rm in}=1\%$ and $A_{\rm in}=0$, at different scales.}
\label{pdfic}
\end{figure}

\begin{figure}[htb]
\includegraphics[width=0.5\textwidth]{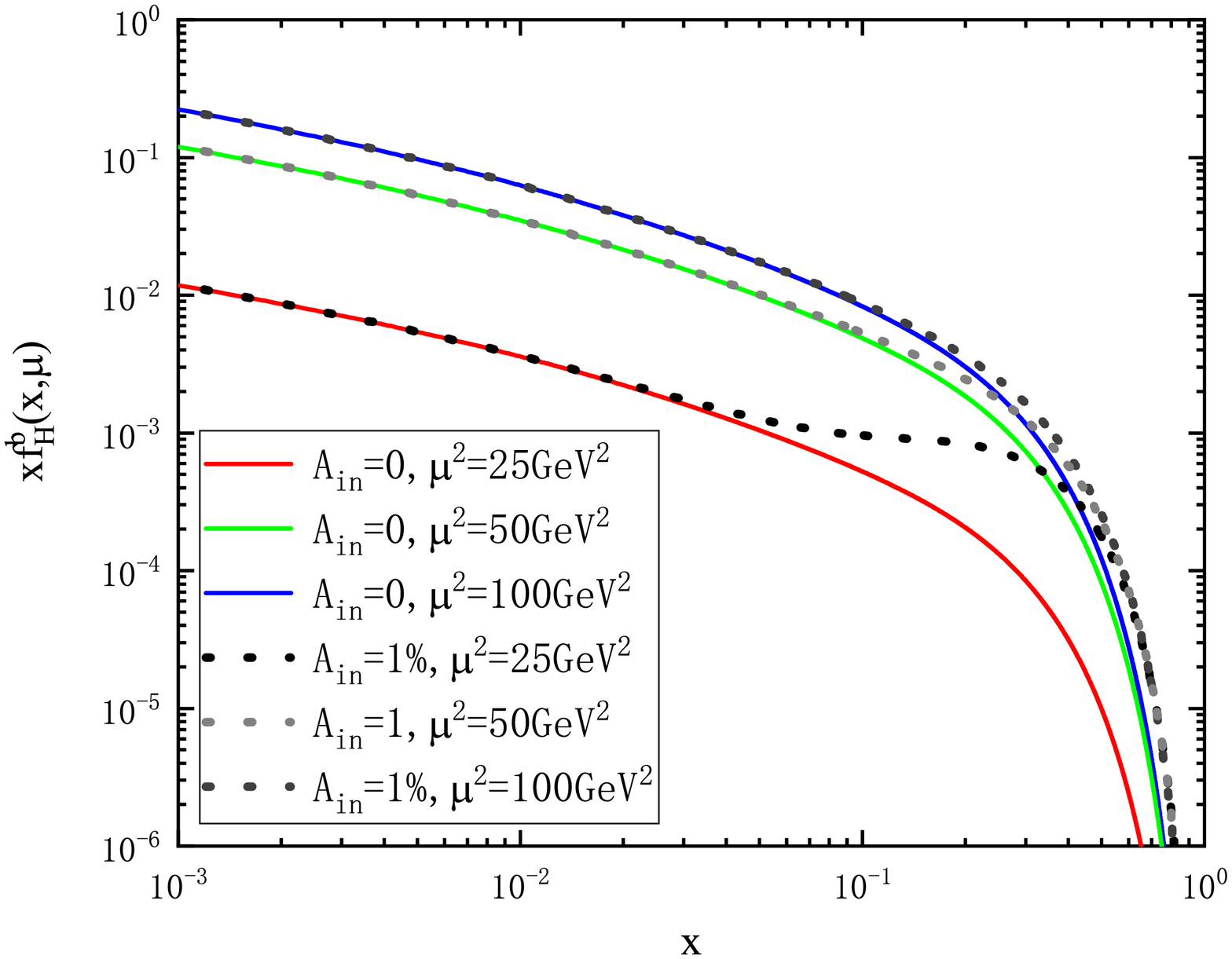}
\caption{The charm and bottom PDFs with and without intrinsic heavy-quark component, $A_{\rm in}=1\%$ and $A_{\rm in}=0$, at different scales.}
\label{pdfib}
\end{figure}

\begin{figure}[htb]
\includegraphics[width=0.5\textwidth]{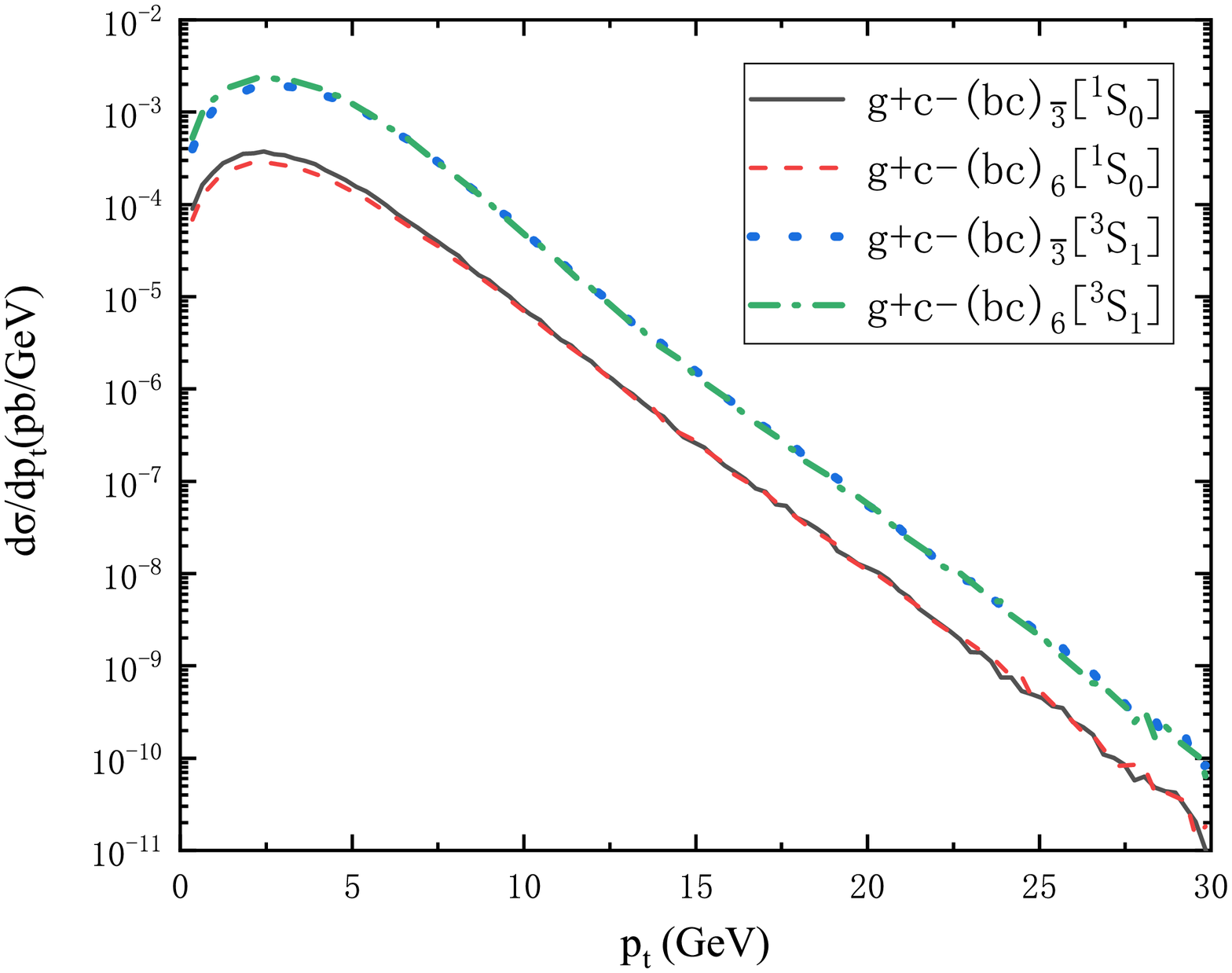}
\caption{The $p_t$ distributions of $\Xi_{bc}$ via the $(g+c)$ mechanism at the After@LHC with intrinsic charm component as $A_{\rm in}=1\%$, in which no $y$ cut has been applied. The contributions from different intermediate $(bc)$ states are presented explicitly.}
\label{pt2gc}
\end{figure}

\begin{figure}[htb]
\includegraphics[width=0.5\textwidth]{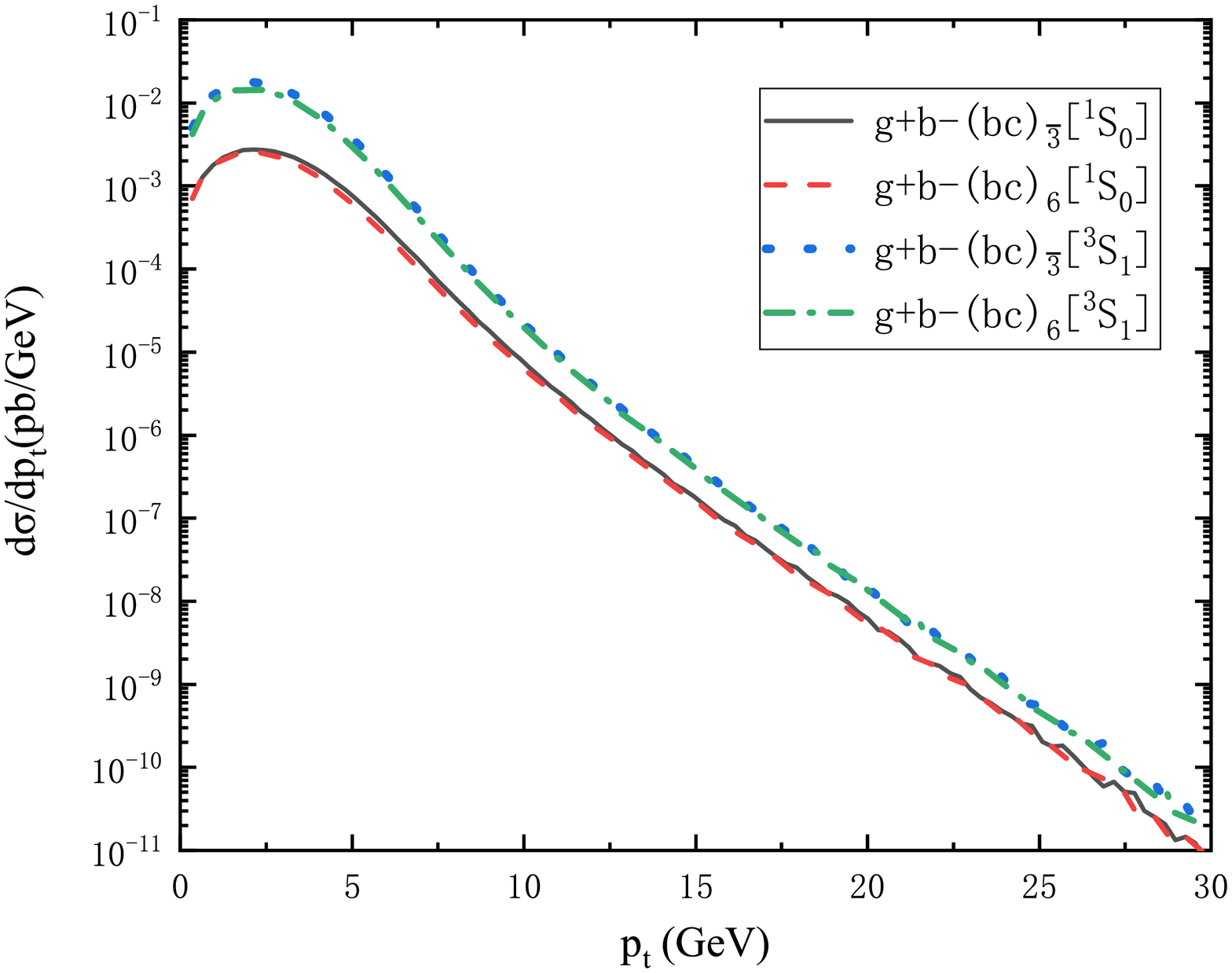}
\caption{The $p_t$ distributions of $\Xi_{bc}$ via the $(g+b)$ mechanism at the After@LHC with the intrinsic charm component as $A_{\rm in}=1\%$, in which no $y$ cut has been applied, and the contributions from different intermediate $(bc)$ states are presented explicitly.}
\label{pt2gb}
\end{figure}

\begin{figure}[htb]
\includegraphics[width=0.5\textwidth]{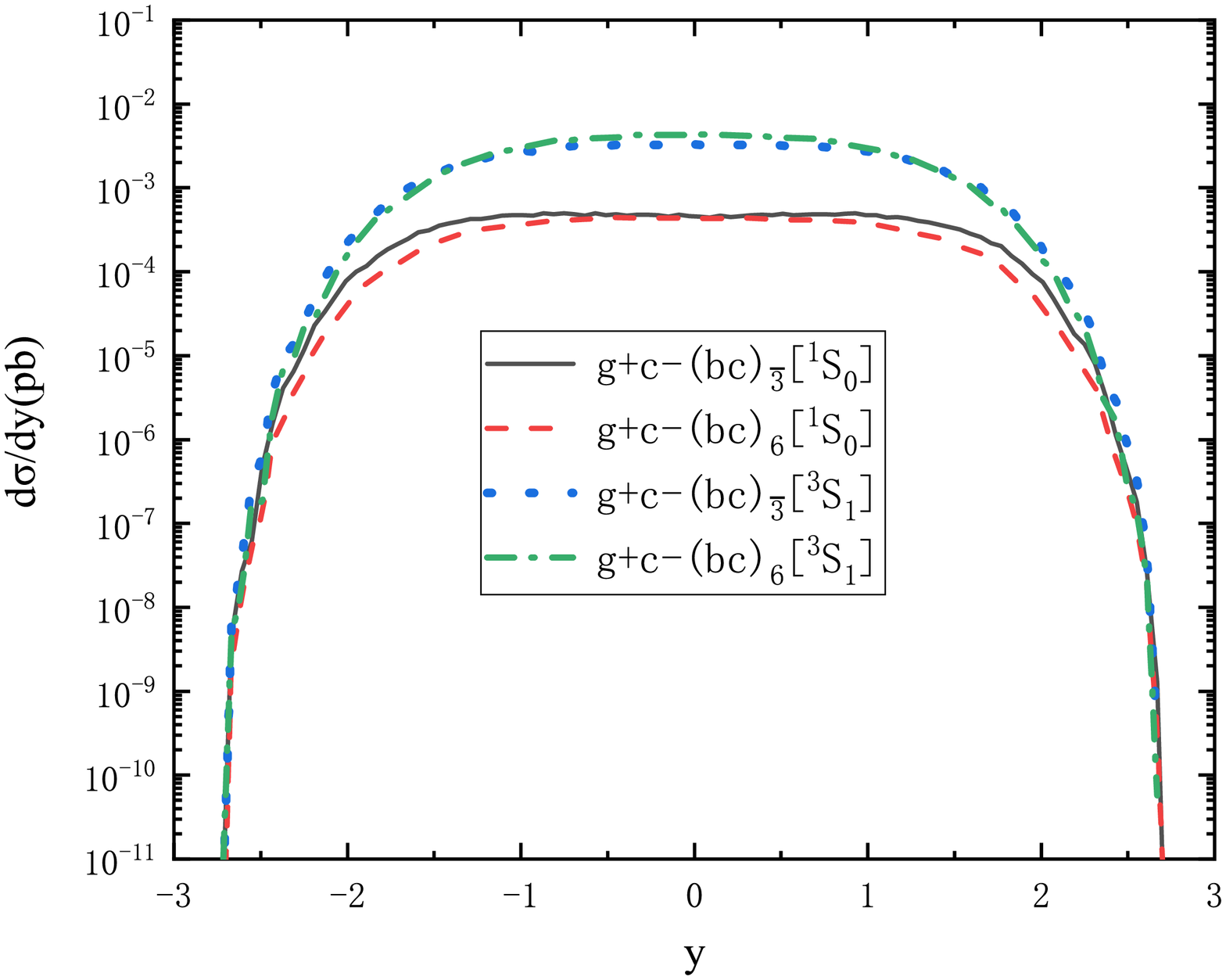}
\caption{The $y$ distributions of $\Xi_{bc}$ via the $(g+c)$ mechanism at the After@LHC with the intrinsic charm component as $A_{\rm in}=1\%$, in which the cut $p_t>0.2\;\rm GeV$ has been taken, and the contributions from different intermediate $(bc)$ states are presented explicitly.}
\label{rap2gc}
\end{figure}

\begin{figure}[htb]
\includegraphics[width=0.5\textwidth]{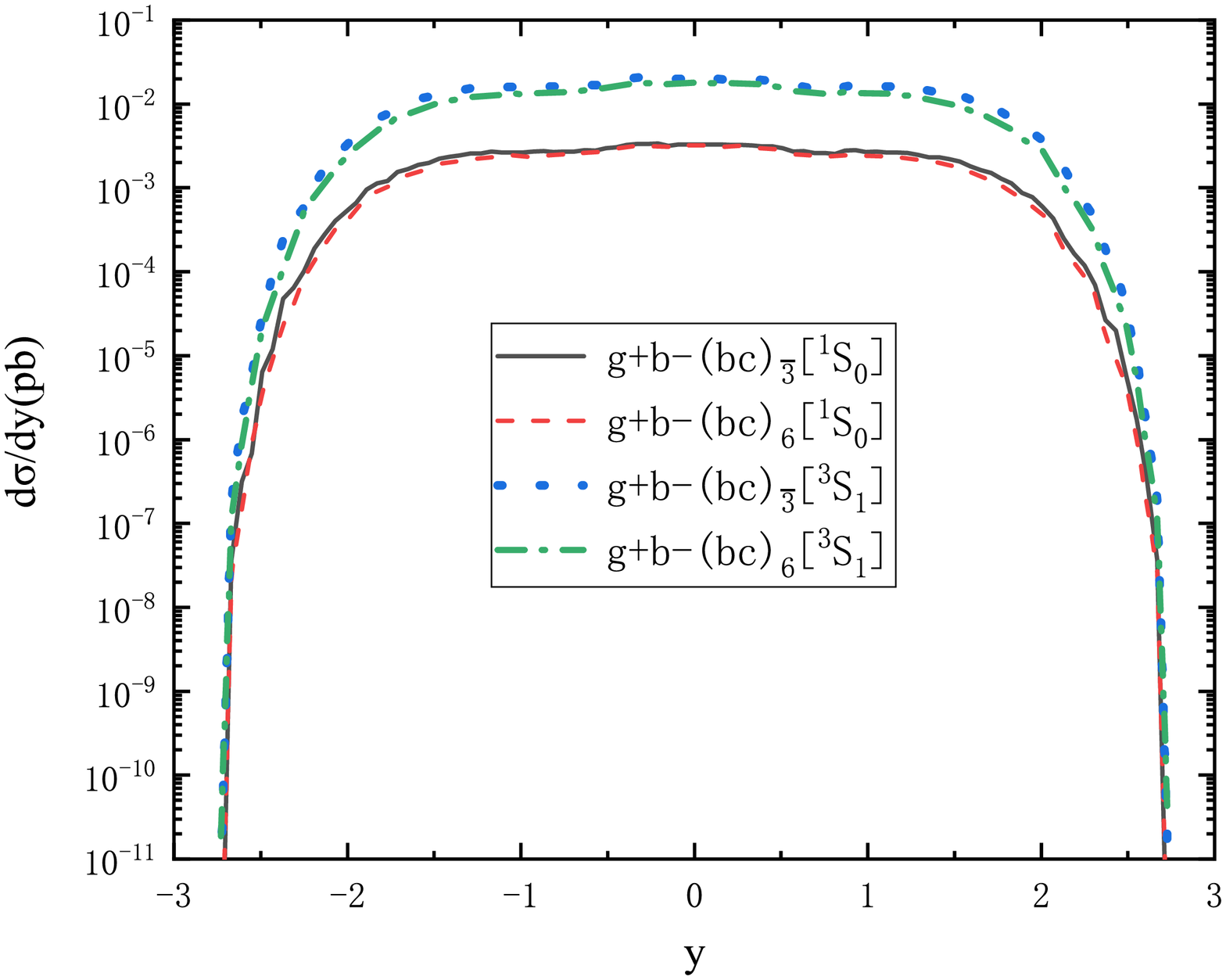}
\caption{The $y$ distributions of $\Xi_{bc}$ via the $(g+b)$ mechanism at the After@LHC with the intrinsic charm component as $A_{\rm in}=1\%$, in which the cut $p_t>0.2\;\rm GeV$ has been taken, and the contributions from different intermediate $(bc)$ states are presented explicitly.}
\label{rap2gb}
\end{figure}

To show how the differential distributions of the $\Xi_{bc}$ production via the $(g+c)$ and $(g+b)$ mechanisms are affected by the intrinsic heavy quarks, we present the $p_t$ and $y$ distributions for $A_{\rm in}=0, 0.3\%, 1\%$ in Figs.\ref{ptgc} and \ref{rapgc}, respectively. Here, the contributions from the $(bc)_{\bar{{\bf 3}}}[^1S_0]$, $(bc)_{{\bf 6}}[^1S_0]$, $(bc)_{\bar{{\bf 3}}}[^3S_1]$ and $(bc)_{{\bf 6}}[^3S_1]$ intermediate states have been summed up. The shapes of the $p_t$ distribution curves for the different $A_{\rm in}$ values are similar. However, it is clear that the normalization of the $p_t$ distribution is changed with the variation of $A_{\rm in}$. The change of the $y$ distribution is more significant with the variation of $A_{\rm in}$. For example, the shape and the normalization of the $y$ distribution change significantly with the increase of $A_{\rm in}$. The changes of these distributions are sizable, which are consistent with the total cross sections in Table \ref{tb.section}, then the measured differential distributions could be potentially adopted to confirm the intrinsic heavy quarks in a proton.

To illustrate how the intrinsic heavy-quark component affects the charm and bottom PDFs, we present the charm and bottom PDFs under several typical scales in Fig.\ref{pdfic} and \ref{pdfib}. The figures show that the charm and bottom PDFs are significantly enhanced in the region of $x \gtrsim 0.2$ after including the intrinsic heavy-quark component. This explains the strong enhancement of the intrinsic heavy quarks to the $\Xi_{bc}$ production via the $(g+c)$ and $(g+b)$ mechanisms at the After@LHC. Therefore, the intrinsic heavy quarks, if exist in hadrons, will play an important role in the hadronic production of $\Xi_{bc}$.

To see the contributions from different intermediate $(bc)$ states via the $(g+c)$ and $(g+b)$ mechanisms, we present the differential $p_t$ and $y$ distributions for the $\Xi_{bc}$ production via different intermediate states, i.e., $(bc)_{\bar{{\bf 3}}}[^1S_0]$, $(bc)_{{\bf 6}}[^1S_0]$, $(bc)_{\bar{{\bf 3}}}[^3S_1]$ and $(bc)_{{\bf 6}}[^3S_1]$, at the After@LHC in Figs.\ref{pt2gc}, \ref{pt2gb}, \ref{rap2gc} and \ref{rap2gb}, respectively. For the $(g+c)$ mechanism, the largest contribution from the $(bc)_{{\bf 6}}[^3S_1]$ channel, while for the $(g+b)$ mechanism, the largest contribution from the $(bc)_{\bar{{\bf 3}}}[^3S_1]$ channel. For both the $(g+c)$ and $(g+b)$ mechanisms, the curves of the $(bc)_{{\bf 6}}[^3S_1]$ and the $(bc)_{\bar{{\bf 3}}}[^3S_1]$ channels are very close.

\subsection{$\Xi_{bc}$ production via the three mechanisms}

In the previous two subsections, we have analyzed the cross sections from different mechanisms respectively. Those results can help us understand the production mechanism of $\Xi_{bc}$ at the After@LHC. In order to compare with experiments, the contributions from different mechanisms should be summed up. In this subsection, we shall analyze the cross sections for the $\Xi_{bc}$ production, where the contributions from the $(g+g)$, $(g+c)$ and $(g+b)$ mechanisms are summed up.

\begin{figure}[htb]
\includegraphics[width=0.5\textwidth]{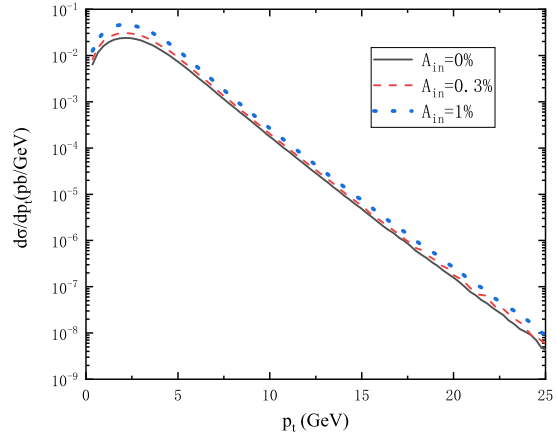}
\caption{Comparison of the $p_t$ distributions for the $\Xi_{bc}$ production with and without intrinsic heavy quarks, $A_{\rm in}=1\%$, $A_{\rm in}=0.3\%$ and $A_{\rm in}=0$ at the After@LHC. Here, the contributions from the $(g+g)$, $(g+c)$ and $(g+b)$ mechanisms have been summed up.}  \label{pttotaltu}
\end{figure}

\begin{figure}[htb]
\includegraphics[width=0.5\textwidth]{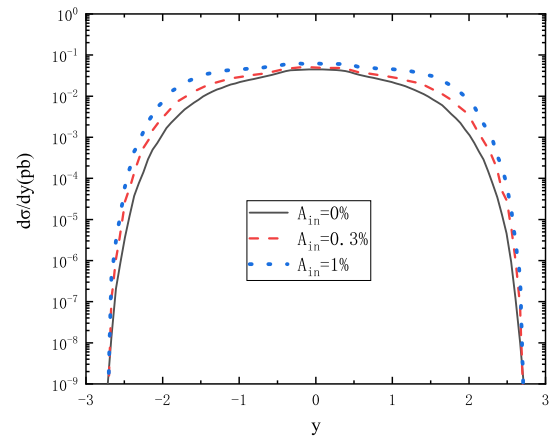}
\caption{Comparison of the $y$ distributions for the $\Xi_{bc}$ production with and without intrinsic heavy quarks, $A_{\rm in}=1\%$, $A_{\rm in}=0.3\%$ and $A_{\rm in}=0$ at the After@LHC. Here, the contributions from the $(g+g)$, $(g+c)$ and $(g+b)$ mechanisms have been summed up.}  \label{raptotaltu}
\end{figure}

In Figs.\ref{pttotaltu} and \ref{raptotaltu}, the differential $p_t$ and $y$ distributions with and without intrinsic heavy quarks are shown, where the contributions from the $(g+g)$, $(g+c)$ and $(g+b)$ mechanisms have been summed up. The figures show that the differential $p_t$ and $y$ distributions are significantly changed after including the intrinsic heavy quarks. This is because the $(g+c)$ and $(g+b)$ mechanisms have important contributions to the cross sections of the $\Xi_{bc}$ production, and the contributions from the $(g+c)$ and $(g+b)$ mechanisms are sensitive to the intrinsic heavy quarks.

\begin{figure}[htb]
\includegraphics[width=0.5\textwidth]{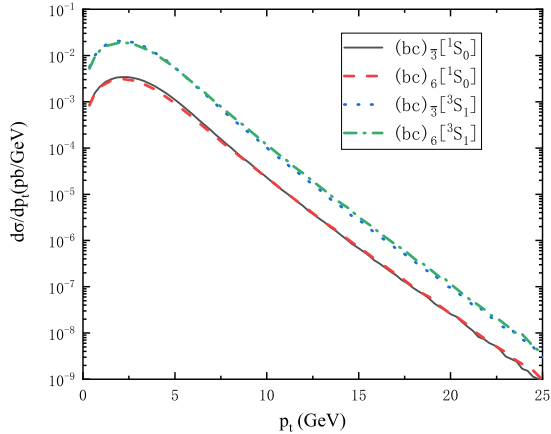}
\caption{The $p_t$ distributions of $\Xi_{bc}$ at the After@LHC with $A_{\rm in}=1\%$, in which no $y$ cut has been applied. The contributions from the $(g+g)$, $(g+c)$ and $(g+b)$ mechanisms have been summed up, while the contributions from different intermediate $(bc)$ states are presented explicitly.}
\label{ptz}
\end{figure}

\begin{figure}[htb]
\includegraphics[width=0.5\textwidth]{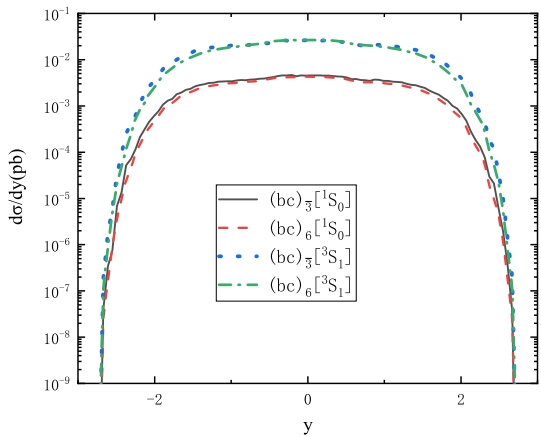}
\caption{The $y$ distributions of $\Xi_{bc}$ at the After@LHC with $A_{\rm in}=1\%$, in which $p_t>0.2\;\rm GeV$ has been applied. The contributions from the $(g+g)$, $(g+c)$ and $(g+b)$ mechanisms have been summed up, while the contributions from different intermediate $(bc)$ states are presented explicitly.}
\label{rapz}
\end{figure}

In Figs.\ref{ptz} and \ref{rapz}, the contributions from different intermediate $(bc)$ states are shown explicitly. The figures show that the curves of different intermediate $(bc)$ states are similar in shape. The dominant contributions to the differential cross sections come from the $(bc)_{\bar{{\bf 3}}}[^3S_1]$ and $(bc)_{{\bf 6}}[^3S_1]$ channels, and they are very close.

\begin{center}
\begin{table}[htb]
\begin{tabular}{|c|c|c|c|c|}
\hline
- &$p_t\ge2$GeV &$p_t\ge4$GeV &$p_t\ge6$GeV &$p_t\ge8$GeV \\
\hline
$\sigma_{(bc)_{\bar{{\bf 3}}}[^1S_0]}$ &6.76 & 1.60&0.38 &0.09 \\
\hline
$\sigma_{(bc)_{{\bf 6}}[^1S_0]}$       &5.43 &1.40 &0.36 &0.09 \\
\hline
$\sigma_{(bc)_{\bar{{\bf 3}}}[^3S_1]}$ &32.60 &7.22 &1.69 &0.40  \\
\hline
$\sigma_{(bc)_{{\bf 6}}[^3S_1]}$       &31.68 &8.14 &2.01 &0.48  \\
\hline
Total                                  &76.47 &18.36 &4.45 &1.06  \\
\hline
\end{tabular}
\caption{Cross sections (in unit pb) of the $\Xi_{bc}$ production at the After@LHC under different $p_t$ cuts, where the contributions from the $(g+g)$, $(g+c)$ and $(g+b)$ mechanisms have been summed up and the probability of the intrinsic charm has been set as $A_{\rm{in}}=1\%$.}
\label{ptcgg1}
\end{table}
\end{center}

\begin{table}[htb]
\begin{tabular}{|c|c|c|c|}
\hline
- & $|y|<1$ & $|y|<2$ & $|y|<3$ \\
\hline
$\sigma_{(bc)_{\bar{{\bf 3}}}[^1S_0]}$ &8.08&12.86 &13.06  \\
\hline
$\sigma_{(bc)_{{\bf 6}}[^1S_0]}$ &7.45 &11.45  &11.61   \\
\hline
$\sigma_{(bc)_{\bar{{\bf 3}}}[^3S_1]}$ &46.14 &73.65 &74.93    \\
\hline
$\sigma_{(bc)_{{\bf 6}}[^3S_1]}$ &46.71 &69.57  &70.53    \\
\hline
Total                             &108.38 &167.53 &170.14   \\
\hline
\end{tabular}
\caption{Cross sections (in unit pb) of the $\Xi_{bc}$ production at the After@LHC under different $y$ cuts, where $A_{\rm{in}}=1\%$, $p_t>0.2\;\rm GeV$ and the contributions from the $(g+g)$, $(g+c)$ and $(g+b)$ mechanisms have been summed up.}
\label{ycgg1}
\end{table}

Considering that the future experiments may adopt kinematic cuts other than the default values (i.e. $p_t>0.2{\rm GeV}$ and no $y$ cut) used in this paper, we present the cross sections under various kinematic cuts in Tables~\ref{ptcgg1} and \ref{ycgg1}, where we have set $A_{\rm in}=1\%$ for the intrinsic heavy quarks. The results show that nearly $89\%$ $\Xi_{bc}$ events to be generated in small $p_t$ region, $p_t\in[0.2,4]\;{\rm GeV}$, and about $64\%$ $\Xi_{bc}$ events for $|y|\le1$.

\subsection{Theoretical uncertainties for the $\Xi_{bc}$ production}

In this subsection, we will discuss the theoretical uncertainties for the $\Xi_{bc}$ production at the After@LHC. The main uncertainty sources include the LDMEs, the charm quark mass, the bottom quark mass, and the renormalization/factorization scale. Up to now, we do not have accurate values for the color ${\bf 6}$ LDMEs, which can be extracted from the future experiments for the $\Xi_{bc}$ production. In this paper, we estimate them by taking them as the same values as the color ${\bf\bar{3}}$ LDMEs. Fortunately, these LDMEs appear as overall factors, so we can easily improve the numerical results when we have accurate values for these LDMEs. Therefore, we will not consider the uncertainties caused by the LDMEs but concentrate our attention on the uncertainties caused by the heavy quark masses and the renormalization/factorization scale. For clarity, when we discuss the uncertainty from one parameter, other input parameters will be kept to be their central values. For convenience, we set $A_{\rm{in}}=1\%$ throughout this subsection.

\begin{center}
\begin{table}[htb]
\begin{tabular}{|c|c|c|c|}
\hline
$m_c$ (GeV)& ~~1.7~~ & ~~1.8~~ & ~~1.9~~\\
\hline
$g+g \to (bc)_{\bar{{\bf 3}}}[^1S_0]$ & ~$1.90$ & ~$1.48$ & ~$1.17$ \\
\hline
$g+g \to (bc)_{{\bf 6}}[^1S_0]$       & ~$1.82$ & ~$1.37$ & ~$1.05$\\
\hline
$g+g \to (bc)_{\bar{{\bf 3}}}[^3S_1]$ & ~$6.85$ & ~$5.29$ & ~$4.13$ \\
\hline
$g+g \to (bc)_{{\bf 6}}[^3S_1]$       & $12.36$ & ~$9.67$ & ~$7.64$\\
\hline
$g+c \to (bc)_{\bar{{\bf 3}}}[^1S_0]$ & ~$1.79$ & ~$1.59$ & ~$1.41$\\
\hline
$g+c \to (bc)_{{\bf 6}}[^1S_0]$       & ~$1.42$ & ~$1.27$ & ~$1.14$ \\
\hline
$g+c \to (bc)_{\bar{{\bf 3}}}[^3S_1]$ & $10.00$ & ~$9.14$ & ~$8.37$  \\
\hline
$g+c \to (bc)_{{\bf 6}}[^3S_1]$       & $11.70$ & $10.67$ & ~$9.75$\\
\hline
$g+b \to (bc)_{\bar{{\bf 3}}}[^1S_0]$ & $13.57$ & $10.03$ & ~$7.54$ \\
\hline
$g+b \to (bc)_{{\bf 6}}[^1S_0]$       & $11.97$ & ~$8.97$ & ~$6.84$ \\
\hline
$g+b \to (bc)_{\bar{{\bf 3}}}[^3S_1]$ & $79.29$ & $60.51$ & $46.92$  \\
\hline
$g+b \to (bc)_{{\bf 6}}[^3S_1]$       & $65.00$ & $50.17$ & $39.34$ \\
\hline
\end{tabular}
\caption{Cross sections (in unit pb) for the $\Xi_{bc}$ production at the After@LHC with a variation of $m_c=1.8\pm0.1\,\rm GeV$, where $p_t>0.2$ GeV and $A_{\rm{in}}=1\%$.}
\label{mass1}
\end{table}
\end{center}

\begin{center}
\begin{table}[htb]
\begin{tabular}{|c|c|c|c|}
\hline
$m_b$ (GeV)& ~~4.9~~ & ~~5.1~~ & ~~5.3~~\\
\hline
$g+g \to (bc)_{\bar{{\bf 3}}}[^1S_0]$ & ~$1.85$ & ~$1.48$ & ~$1.19$\\
\hline
$g+g \to (bc)_{{\bf 6}}[^1S_0]$       & ~$1.68$ & ~$1.37$ & ~$1.13$\\
\hline
$g+g \to (bc)_{\bar{{\bf 3}}}[^3S_1]$ & ~$6.62$ & ~$5.29$ & ~$4.25$  \\
\hline
$g+g \to (bc)_{{\bf 6}}[^3S_1]$       & $12.14$ & ~$9.67$ & ~$7.74$\\
\hline
$g+c \to (bc)_{\bar{{\bf 3}}}[^1S_0]$ & ~$1.99$ & ~$1.59$ & ~$1.27$  \\
\hline
$g+c \to (bc)_{{\bf 6}}[^1S_0]$       & ~$1.60$ & ~$1.27$ & ~$1.02$ \\
\hline
$g+c \to (bc)_{\bar{{\bf 3}}}[^3S_1]$ & $11.69$ & ~$9.14$ & ~$7.19$ \\
\hline
$g+c \to (bc)_{{\bf 6}}[^3S_1]$       & $13.57$ & $10.67$ & ~$8.45$   \\
\hline
$g+b \to (bc)_{\bar{{\bf 3}}}[^1S_0]$ & ~$9.05$ & $10.03$ & $10.75$\\
\hline
$g+b \to (bc)_{{\bf 6}}[^1S_0]$       & ~$8.11$ & ~$8.97$ & ~$9.59$\\
\hline
$g+b \to (bc)_{\bar{{\bf 3}}}[^3S_1]$ & $55.46$ & $60.51$ & $63.88$ \\
\hline
$g+b \to (bc)_{{\bf 6}}[^3S_1]$       & $46.10$ & $50.17$ & $52.69$  \\
\hline
\end{tabular}
\caption{Cross sections (in unit pb) for the $\Xi_{bc}$ production at the After@LHC with a variation of $m_b=5.1\pm0.2\,\rm GeV$, where $p_t>0.2$ GeV and $A_{\rm{in}}=1\%$.}
\label{mass2}
\end{table}
\end{center}

We first consider the uncertainties caused by the charm and bottom quark masses. We estimate them by taking $m_c=1.8\pm0.1\,\rm GeV$ and $m_b=5.1\pm0.2\,\rm GeV$. The uncertainties caused by the charm and bottom quark masses are presented in Tables~\ref{mass1} and \ref{mass2}, respectively. From the tables, we can see that the cross sections for the $(g+g)$ and $(g+b)$ mechanisms are more sensitive to the charm quark mass than that for the $(g+c)$ mechanism. For example, when the charm quark mass is increased by $0.1\,{\rm GeV}$, the cross sections for the $(g+g)$ and $(g+b)$ mechanisms are decreased by $21\%$ and $22\%$ respectively, while the cross section for the $(g+b)$ mechanism is only decreased by $9\%$. The cross sections for the $(g+g)$ and $(g+c)$ mechanisms are more sensitive to the bottom quark mass than that for the $(g+b)$ mechanism. When the bottom quark mass is increased by $0.2\,{\rm GeV}$, the cross sections for the $(g+g)$ and $(g+c)$ mechanisms are decreased by $20\%$ and $21\%$ respectively, while the cross section for the $(g+b)$ mechanism is increased by $6\%$.

Then we consider the uncertainties caused by the renormalization and factorization scales. For simplicity, we also set the renormalization and factorization scales to be the same, i.e., $\mu_F=\mu_R=\mu$, and take them as the three typical energy scales ($M_{\Xi_{bc}}$, $m_T$, and the center-of-mass energy of the
parton subprocess $\sqrt{\hat{s}}$) involved in the production process to estimate the uncertainties. The uncertainties caused by the renormalization/factorization scale are presented in Table \ref{scale1}.

\begin{center}
\begin{table}[htb]
\begin{tabular}{|c|c|c|c|}
\hline
$\mu$& ~~$M_{\Xi_{bc}}$~~ &  ~~$m_T$~~  &  ~~$\sqrt{\hat{s}}$~~ \\
\hline
$g+g \to (bc)_{\bar{{\bf 3}}}[^1S_0]$ & ~$1.84$ & ~$1.48$ & ~$0.39$ \\
\hline
$g+g \to (bc)_{{\bf 6}}[^1S_0]$       & ~$1.74$ & ~$1.37$ & ~$0.36$\\
\hline
$g+g \to (bc)_{\bar{{\bf 3}}}[^3S_1]$ & ~$6.56$ & ~$5.29$ & ~$1.32$ \\
\hline
$g+g \to (bc)_{{\bf 6}}[^3S_1]$       & $11.92$ & ~$9.67$ & ~$2.55$  \\
\hline
$g+c \to (bc)_{\bar{{\bf 3}}}[^1S_0]$ & ~$1.74$ & ~$1.59$ & ~$0.85$ \\
\hline
$g+c \to (bc)_{{\bf 6}}[^1S_0]$       & ~$1.40$ & ~$1.27$ & ~$0.73$ \\
\hline
$g+c \to (bc)_{\bar{{\bf 3}}}[^3S_1]$ & $10.19$ & ~$9.14$ & ~$5.26$   \\
\hline
$g+c \to (bc)_{{\bf 6}}[^3S_1]$       & $11.77$ & $10.67$ & ~$6.34$  \\
\hline
$g+b \to (bc)_{\bar{{\bf 3}}}[^1S_0]$ & $10.29$ & $10.03$ & ~$8.85$ \\
\hline
$g+b \to (bc)_{{\bf 6}}[^1S_0]$       & ~$9.18$ & ~$8.97$ & ~$8.15$ \\
\hline
$g+b \to (bc)_{\bar{{\bf 3}}}[^3S_1]$ & $61.75$ & $60.51$ & $54.16$   \\
\hline
$g+b \to (bc)_{{\bf 6}}[^3S_1]$       & $51.12$ & $50.17$ & $45.67$  \\
\hline
\end{tabular}
\caption{Cross sections (in unit pb) for the $\Xi_{bc}$ production at the After@LHC under the different choices of the renormalization/factorization scale (i.e. $\mu=M_{\Xi_{bc}}$, $m_T$ and $\sqrt{\hat{s}}$), where $p_t>0.2$ GeV and $A_{\rm{in}}=1\%$.}
\label{scale1}
\end{table}
\end{center}

To obtain total uncertainties from the heavy quark masses and the renormalization/factorization scale, we adding the uncertainties in quadrature. Then we obtain
\begin{eqnarray}
\sigma_{g+g \to (bc)_{\bar{{\bf 3}}}[^1S_0]} &=& 1.48^{+0.66}_{-1.17}\;{\rm pb}, \nonumber\\
\sigma_{g+g \to (bc)_{\bf 6}[^1S_0]} &=& 1.37^{+0.66}_{-1.08}\;{\rm pb}, \nonumber\\
\sigma_{g+g \to (bc)_{\bar{\bf 3}}[^3S_1]} &=& 5.29^{+2.41}_{-4.26}\;{\rm pb}, \nonumber\\
\sigma_{g+g \to (bc)_{{\bf 6}}[^3S_1]} &=& 9.67^{+4.29}_{-7.70}\;{\rm pb}, \nonumber\\
\sigma_{g+c \to (bc)_{\bar{{\bf 3}}}[^1S_0]} &=& 1.59^{+0.47}_{-0.82}\;{\rm pb}, \nonumber\\
\sigma_{g+c \to (bc)_{\bf 6}[^1S_0]} &=& 1.27^{+0.39}_{-0.62}\;{\rm pb}, \nonumber\\
\sigma_{g+c \to (bc)_{\bar{\bf 3}}[^3S_1]} &=& 9.14^{+2.89}_{-4.41}\;{\rm pb}, \nonumber\\
\sigma_{g+c \to (bc)_{{\bf 6}}[^3S_1]} &=& 10.67^{+3.27}_{-4.95}\;{\rm pb}, \nonumber\\
\sigma_{g+b \to (bc)_{\bar{{\bf 3}}}[^1S_0]} &=& 10.03^{+3.62}_{-2.92}\;{\rm pb}, \nonumber\\
\sigma_{g+b \to (bc)_{\bf 6}[^1S_0]} &=& 8.97^{+3.07}_{-2.44}\;{\rm pb}, \nonumber\\
\sigma_{g+b \to (bc)_{\bar{\bf 3}}[^3S_1]} &=& 60.51^{+19.12}_{-15.83}\;{\rm pb}, \nonumber\\
\sigma_{g+b \to (bc)_{{\bf 6}}[^3S_1]} &=& 50.17^{+15.07}_{-12.41}\;{\rm pb}.
\end{eqnarray}
Adding up the contributions from the different intermediate $(bc)$ states, we obtain
\begin{eqnarray}
\sigma_{g+g} &=& 17.81^{+8.02}_{-14.21}\;{\rm pb}, \nonumber\\
\sigma_{g+c} &=& 22.67^{+7.02}_{-10.80}\;{\rm pb}, \nonumber\\
\sigma_{g+b} &=& 129.68^{+40.88}_{-33.60}\;{\rm pb},
\end{eqnarray}
and
\begin{eqnarray}
\sigma_{\rm Total} &=& 170.16^{+55.92}_{-58.61}\;{\rm pb}.
\end{eqnarray}

\section{Conclusions}
\label{summary}

In the present paper, we have studied the hadronic production of the $\Xi_{bc}$ baryon at the suggested fixed-target experiment After@LHC. The integrated cross sections and the differential distributions ($d\sigma/dp_t$ and $d\sigma/dy$) for the $\Xi_{bc}$ production are obtained, and the main theoretical uncertainties for the cross sections are analyzed.

For the initial partons, in addition to the $(g+g)$ fusion mechanism, the $(g+c)$ and $(g+b)$ collision mechanisms are also considered. It is found that the $(g+c)$ and $(g+b)$ mechanisms give sizable contributions to the $\Xi_{bc}$ production, and the $(g+b)$ mechanism dominates the production. For the initial heavy quarks, both the extrinsic and the intrinsic components are considered. The results show that the intrinsic heavy quarks can have significant impact on the $\Xi_{bc}$ production. If setting the probability for the intrinsic charm, $A_{\rm in}=1\%$, the cross section for the $(g+c)$ mechanism will be enhanced by about $80\%$; and if further setting the probability for the intrinsic bottom to be one order smaller than the probability of intrinsic charm, the cross section for the $(g+b)$ mechanism will be enhanced by about $114\%$. Thus, the intrinsic heavy quarks play important roles in the $\Xi_{bc}$ baryon at the After@LHC.

If the integrated luminosity of the After@LHC can be up to $2\,{\rm fb}^{-1}$ per year, there about $3.40\times10^5$ $\Xi_{bc}$ events to be produced per year. Therefore, the $\Xi_{bc}$ may be observed at the After@LHC, and the intrinsic heavy quarks, may be tested and studied through studying the $\Xi_{bc}$ production at the After@LHC.

\hspace{2cm}

{\bf Acknowledgements}: This work was supported in part by the Natural Science Foundation of China under Grants No.12005028, No.12175025 and No.12147102, by the China Postdoctoral Science Foundation under Grant No.2021M693743, by the Fundamental Research Funds for the Central Universities under Grant No.2020CQJQY-Z003, by the Chongqing Natural Science Foundation under Grant No.CSTB2022NSCQ-MSX0415, and by the Chongqing Graduate Research and Innovation Foundation under Grant No.ydstd1912.


\end{document}